\documentclass{llncs}

\usepackage{stmaryrd}
\usepackage{bbm}
\usepackage{epsf}

\begin{document}

\newcommand{\al}{\alpha}
\newcommand{\mb}{\mbox}
\newcommand{\R}{\mathbbmss{R}}
\newcommand{\N}{\mathbbmss{N}}
\newcommand{\Q}{\mathbbmss{Q}}
\newcommand{\Z}{\mathbbmss{Z}}

\title{The Three Gap theorem\\
(Steinhaus conjecture)}

\author{Micaela Mayero\thanks{{\sf Micaela.Mayero@inria.fr}, 
{\sf http://coq.inria.fr/\~{}mayero/}}}

\institute{INRIA-Rocquencourt\thanks{Projet {\sf Coq}, INRIA-Rocquencourt, 
domaine de Voluceau, B.P. 105, 78153 Le Chesnay Cedex, France.}}

\maketitle

\begin{abstract}
We deal with the distribution of $N$ points placed consecutively 
around the circle by a fixed angle of $\al{}$. From the proof of Tony 
van Ravenstein \cite{RAV88}, we propose a detailed proof of the 
Steinhaus conjecture whose result is the following: the $N$ points
partition the circle into gaps of at most three different lengths.\\
We study the mathematical notions required for the proof of 
this theorem revealed during a formal proof carried out in {\sf Coq}.
\end{abstract}

\section*{Introduction}

Originally, the three gap theorem was the conjecture of H. Steinhaus.
Subsequently, several proofs were proposed by \cite{SOS57} 
\cite{SOS58} \cite{SWI58} \cite{SUR58} \cite{HAL65} \cite{SLA67} 
\cite{RAV88}. The proof proposed in this paper is a presentation 
of the proof completely fomalized in the {\sf Coq} proof asssistance
system \cite{MAY99}. This formal proof is based on Tony van 
Ravenstein's \cite{RAV88}.

This kind of demostration, which involves geometrical intuition,
is a real challenge for proof assistance systems. That is what 
motivated our work. Therefore, the interest of such an approach is to 
understand, by means of an example, if the {\sf Coq} system allows us
to prove a theorem coming from pure mathematics.\\
In addition, this development allowed us to clarify some points of 
the proof and has led to a more detailed proof.

First, we will define the notations and definitions used
for stating and proving this theorem.The second part deals with
different states of the theorem and with the proof itself. Finally,
the last part presents advantages of the formal proof stating the
main differences between our proof and Tony van Ravenstein's
proof.

\section{Notations and definitions}

\subsection{Notations}

\begin{figure}[ht]
\begin{center}
\noindent{}\framebox[10cm][l]
{\parbox{10cm}{
\epsfxsize=10cm
$$
\epsfbox{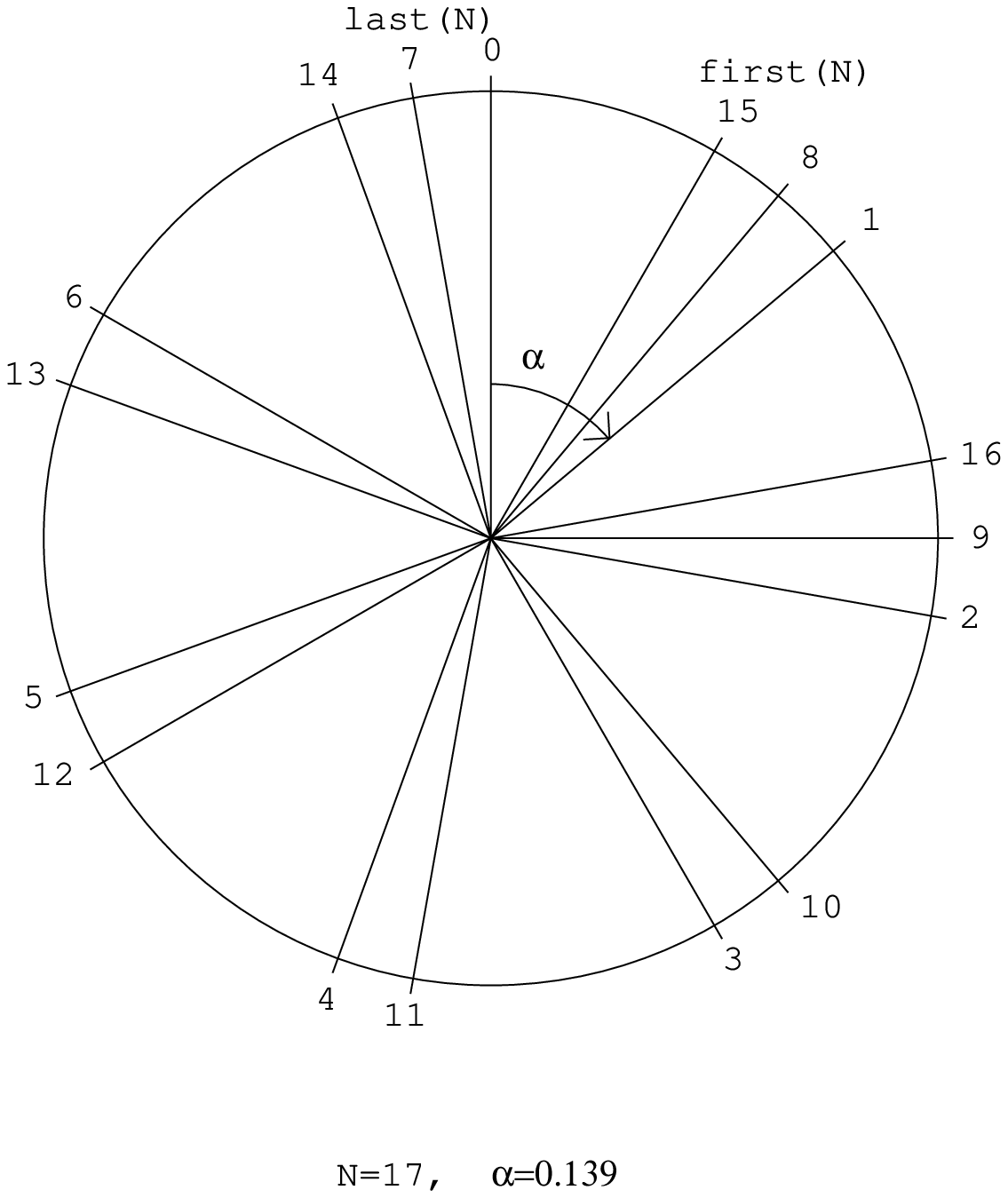}
$$}}
\caption{The three gap theorem}
\label{cercle}
\end{center}
\end{figure}

We can refer to figure \ref{cercle}.

\begin{itemize}
\item $\N{}$ is the natural numbers set.
\item $\R{}$ is the real numbers set.
\item The integer part of a real number r is noted $E(r)$.
\item The fractional part of a real number $r$, ie $r-E(r)$, is 
written $\{r\}$.
\item The first point on the circle is the point $0$.
\item Unless explicitly mentioned, we consider $N$ points 
distributed around the circle. These are numbered from $0$ to $N-1$. 
\item We consider the circle of {\bf unit circumference} and with a 
clockwise orientation.
\item $\al{}$ is counted in turns of the circle (and not in radian); 
then $0\leq{}\al{}<1$.
\item The first point ($\neq{}0$ if $N>$1) on the right of $0$ is 
written $first(N)$. $first(N)$ is a function from the natural numbers 
to the natural numbers.
\item The last point ($\neq{}$0 if $N>$1) before $0$ is written 
$last(N)$. $last(N)$ is a function from the natural numbers to 
the natural numbers. 
\item $n\in{}Circle$ is equivalent to $0\leq{}n<N$.
\end{itemize}

\begin{remark}
The distance from point $0$ to point n is $\{n.\al{}\}$.
\end{remark}

\subsection{Definitions}

The following definitions are valid for all $\al{}$, rational or 
irrational. 

\begin{lemma}[Existence of first]
\label{first}
If $N\geq{}2$ then there exists an integer \linebreak 
$first(N)\in{}\N{}$ s.t. 
$0<first(N)<N$ and $\forall{}m\in{}\N{}$ if  
$0<m<N$ then $\{first(N).\al{}\}\leq{}\{m.\al{}\}$
\end{lemma}

\begin{proof}
\ \\
By induction on $N$.\\
- $N=2$: in this case $first(N)=1$.\\
- Suppose the lemma to be holds for $N$: then 
there exists $first(N)\in{}\N{}$ s.t. \linebreak
$0<first(N)<N$ and  
$\forall{}m\in{}\N{}$ if $0<m<N$ then
$\{first(N).\al{}\}\leq{}\{m.\al{}\}$; \\
let us show that there 
exists $first(N+1)\in{}\N{}$ s.t. 
$0<first(N+1)<N+1$ and  \linebreak
$\forall{}m\in{}\N{}$ if $0<m<N+1$ then  
$\{first(N+1).\al{}\}\leq{}\{m.\al{}\}$.\\
 By cases:

- if $\{first(N).\al{}\}\leq{}\{N.\al{}\}$ then 
$first(N+1)=first(N)$.

- if $\{first(N).\al{}\}>\{N.\al{}\}$ then $first(N)=N$.
\begin{flushright}
$\boxempty{}$
\end{flushright}
\end{proof}

\begin{lemma}[Existence of last]
\label{last}
If $N\geq{}2$ then there exists an integer \linebreak
$last(N)\in{}\N{}$ s.t. 
$0<last(N)<N$ and $\forall{}m\in{}\N{}$ if 
$0<m<N$ then \linebreak
$\{m.\al{}\}\leq{}\{last(N).\al{}\}$
\end{lemma}

\begin{proof}
\ \\
Symmetrical proof with respect to first.
\begin{flushright}
$\boxempty{}$
\end{flushright}
\end{proof}
The successor of a point on the circle ($after$) verifies the 
following property:

\begin{lemma}[Property of points for after]
\label{points}
$\forall{}M\in{}\R{}$ if $0\leq{}M<1$ then we have:
\begin{enumerate}
either 
\item there exists an integer $I\in{}\N{}$ s.t. $0<I<N$ and 
$M<\{I.\al{}\}$ and 
$\forall{}m\in{}\N{}$ \linebreak
if $0\leq{}m<N$ and if $\{m.\al{}\}>M$ then 
$\{m.\al{}\}\geq{}\{I.\al{}\}$\\
or 
\item $\forall{}m\in{}\N{}$ if $0\leq{}m<N$ then 
$0\leq{}\{m.\al{}\}\leq{}M$
\end{enumerate}
\end{lemma}

\begin{proof}
\ \\
By induction on $N$.\\
- $N=1$: $0$ verifies the property.\\
- Suppose lemma to holds for $N$ and prove it for $N+1$: 
the induction hypothesis is the following:\\ 
either\\
1. there exists an integer $I(N)\in{}\N{}$ s.t. $0<I(N)<N$ and 
$M<\{I(N).\al{}\}$ and $\forall{}m\in{}\N{}$ if
$0\leq{}m<N$ and if $\{m.\al{}\}>M$ then 
$\{m.\al{}\}\geq{}\{I(N).\al{}\}$\\
or\\
2. $\forall{}m\in{}\N{}$ if $0\leq{}m<N$ then 
$0\leq{}\{m.\al{}\}\leq{}M$\\
We prove that :\\
either

1. there exists an integer $I(N+1)\in{}\N{}$ s.t. $0<I(N+1)<N+1$ 
and $M<\{I(N+1).\al{}\}$ and 
$\forall{}m\in{}\N{}$ if
$0\leq{}m<N+1$ and if $\{m.\al{}\}>M$ then 
$\{m.\al{}\}\geq{}\{I(N+1).\al{}\}$\\
or

2. $\forall{}m\in{}\N{}$ if $0\leq{}m<N+1$ then 
$0\leq{}\{m.\al{}\}\leq{}M$\\
By cases:

- if $0\leq{}M<\{N.\al{}\}$ we are in case 1 and we continue by 
cases:

\hspace{2em} - if $\{N.\al{}\}<\{I(N).\al{}\}$ then $I(N+1)=N$.

\hspace{2em} - if $\{I(N).\al{}\}\leq{}\{N.\al{}\}$ then 
$I(N+1)=I(N)$.

- if $\{N.\al{}\}\leq{}M<1$ we are in case 2 and the
proof is immediate by induction hypothesis.
\begin{flushright}
$\boxempty{}$
\end{flushright}
\end{proof}

\begin{definition}[after]
\label{after}
For all points n on the circle, the point $after(N,n)$ verifies the
property of points (lemma \ref{points}) for
$M=\{n.\al{}\}$ and is defined such that:\\
if we are in case 1. then $after(N,n)=I$\\
if we are in case 2. then $after(N,n)=0$. 
\end{definition}

\section{Statement and proof of the theorem}

\subsection{Statement}

Statement in natural language:

\begin{theorem}[Intuitive statement] 
\label{theorem1}
Let $N$ points be placed consecutively \linebreak around the circle by an angle
of $\al{}$. Then for all irrational $\al{}$ and natural $N$, the points 
partition the circle into gaps of at most three different lengths. 
\end{theorem}

As shown by theorem \ref{theorem1} the points are numbered in order of
apparition; now, if we do more than one revolution around the circle,
new points appear between the former points. Then, when the last
point $(N-1)$ is placed, it is possible to number them again
consecutively and clockwise. In this new numeration, we use
only the definitions of $first$, $last$ and $after$, from lemmas 
\ref{first} and \ref{last}, and of the definition \ref{after}.

If we set $\parallel$$x$$\parallel$=$min(\{x\},1-\{x\})$,
then the distance of a point $n$ from its successor $after(N,n)$
is given by $\parallel$$after(N,n)-n$$\parallel$. In order to show 
that this function can have at most three values, we show that the
function $(after(N,n)-n)$ can itself have at most three values.\\

So, proving theorem \ref{theorem1} comes to the same thing as 
showing the following mathematical formulation, which we will
prove in the next paragraph.

\begin{theorem}[The three gap theorem]
\label{letheoreme}
$$after(N,m)-m=\left\{\begin{array}{ll}
          first(N) & \mbox{ if } 0\leq{}m<N-first(N)\\
          first(N)-last(N) & \mbox{ if } N-first(N)\leq{}m<last(N)\\
          -last(N) & \mbox{ if } last(N)\leq{}m<N
                   \end{array}
            \right.
$$
\end{theorem}
\begin{remark}
\ \\
\begin{enumerate}
\item This transcription means that the circle of $N$ points 
is divided into \\
$N-first$ gaps of length $\parallel$$first.\al{}\parallel$, \\
$N-last$ gaps of length $\parallel$$last.\al{}\parallel$ and \\
$first+last-N$ gaps of length 
$\parallel$$first.\al{}\parallel$+$\parallel$$last.\al{}\parallel$.
\item Theorem \ref{letheoreme} is true for $\al{}$ rational and 
irrational. Here, however, we present only the proof for $\al{}$
irrational. Indeed, most of the intermediate results are false
for $\al{}$ rational (among other reasons because $first$, $last$ and 
$after$ are no longer functions). Moreover, the theorem is trivially
true for $\al{}$ rational: if $\al{}=p/q$ then the circle may 
include one or two lengths of gap - depending on whether $N<q$ or 
$N=q$.
\end{enumerate}
\end{remark}

\subsection{Proof}
We recall that the proof is detailed for $\al{}$ irrational and
$N\geq{}2$.\\
\begin{lemma}[particular case]
\label{inter1}
If $N=first(N)+last(N)$ 
$$after(N,m)-m=\left\{\begin{array}{ll}
               first(N) & \mbox{ if } 0\leq{}m<last(N)\\
               -last(N) & \mbox{ if } last(N)\leq{}m<N
                   \end{array}
            \right.
$$
\end{lemma}

\begin{proof}
\ \\
\begin{enumerate}
\item Case $0\leq{}m<last(N)$:\\
For $m=0$, by definition of first we have 
$after(N,0)=first(N)$.\\
\\
We want to prove that $m+first(N)$ is the successor of $m$.\\
Let us first show that $m+first(N)$ belongs to the circle of
$N$ points: \linebreak
if $0<m<last(N)$ then \\ 
$0<0+first(N)\leq{}m+first(N)<last(N)+first(N)=N$.\\
\\
Now, let us show that: if $i$ is any point of the circle
($0<i<N$) then we have either $\{i.\al{}\}<\{m.\al{}\}$
or $\{i.\al{}\}>\{(m+first(N)).\al{}\}$. \\
By reductio ad absurdum and by cases: \\
let us suppose that  
$\{m.\al{}\}<\{i.\al{}\}<\{(m+first(N)).\al{}\}$\\
\begin{itemize}
\item  if $i>m$ then 
$0<\{i.\al{}\}-\{m.\al{}\}<\{(m+first(N)).\al{}\}-\{m.\al{}\}$
therefore \\
$0<\{(i-m).\al{}\}<\{first(N).\al{}\}$ which contradicts the
definition of first, since $(i-m)\in{}$Circle because $0<i-m<N$.\\
\item if $i\leq{}m$ then
$\{(m+first(N)).\al{}\}-\{i.\al{}\}<\{(m+first(N)).\al{}\}-
\{m.\al{}\}$ therefore 
$\{(m+first(N)-i).\al{}\}<\{first(N).\al{}\}$ which contradicts
the definition of  first(N), since $(m+first(N)-i)\in{}$Circle 
because $0<m+first(N)-i<N$.\\
\end{itemize}
In these two former cases
$\{(m+first(N)).\al{}\}-\{m.\al{}\}=\{first(N).\al{}\}$ if 
$\{m.\al{}\}<\{(m+first(N)).\al{}\}$. \\
Let us show this by reductio ad absurdum:\\
if $\{(m+first(N)).\al{}\}\leq{}\{m.\al{}\}$ then \\ 
$\{(m+first(N)).\al{}\}-\{first(N).\al{}\}\leq{}\{m.\al{}\}-
\{first(N).\al{}\}$ \\
therefore by definition of $first$ 
$\{m.\al{}\}\leq{}\{m.\al{}\}-\{first(N).\al{}\}$ which is 
absurd because $\{first(N).\al{}\}>0$ for $N\geq{}2$.\\
\item Case $last(N)\leq{}m<N$:\\
For $m=last(N)$, by definition of $last$ we have 
$after(N,last(N))=0$.\\
\\
We want to prove that $m-last(N)$ is the successor of $m$.\\
Let us first show that $m-last(N)$ belongs to the circle of
$N$ points: \linebreak
if $last(N)<m<N$ then \\
$0\leq{}last(N)-last(N)<m-last(N)<N-last(N)<N$ because $last(N)>0$.\\
\\
Now, let us show that: if $i$ any point of the circle
($0<i<N$) then we have either $\{i.\al{}\}<\{m.\al{}\}$ or
$\{i.\al{}\}>\{(m-last(N)).\al{}\}$. \\
By reductio ad absurdum and by cases: \\
let us suppose that 
$\{m.\al{}\}<\{i.\al{}\}<\{(m-last(N)).\al{}\}$\\
\begin{itemize}
\item if $i<m$ then 
$\{m.\al{}\}-\{i.\al{}\}+1>\{m.\al{}\}-\{(m-last(N)).\al{}\}+1$
therefore 
$\{(m-i).\al{}\}>\{last(N).\al{}\}$ which contradicts the
definition of last, since $(m-i)\in{}$Circle because $0<m-i<N$.\\
\item if $m\leq{}i$ then\\
$\{m.\al{}\}-\{(m-last(N)).\al{}\}+1<\{i.\al{}\}-
\{(m-last(N)).\al{}\}+1$ \\
therefore $\{last(N).\al{}\}<\{(i+m-last(N)).\al{}\}$ \\which 
contradicts the definition of last, since 
$(i+m-last(N))\in{}$Circle because $0<i+m-last(N)<N$.\\
\end{itemize}
In these two former cases 
$\{m.\al{}\}-\{(m-last(N)).\al{}\}+1=\{last(N).\al{}\}$ if 
$\{m.\al{}\}<\{(m-last(N)).\al{}\}$. \\
As $\al{}$ is irrationnal and by definition of $last$ we have 
$\{m.\al{}\}<\{last(N).\al{}\}$
therefore 
$\{(m-last(N)).\al{}\}=\{m.\al{}\}-\{last(N).\al{}\}+1$ and 
we have effectively 
$\{m.\al{}\}<\{m.\al{}\}-\{last(N).\al{}\}+1$,
since $\{last(N).\al{}\}<1$.
\end{enumerate}
\end{proof}
\begin{remark}
\label{irrat}
The fact of $\al{}$ is irrational is essential for showing that
\linebreak
$\{m.\al{}\}\neq{}\{last(N).\al{}\}$. Indeed, 
as in this case we have
$m\neq{}last(N)$ therefore $\{m.\al{}\}\neq{}\{last(N).\al{}\}$.
Let us show this by contradiction. \\
In order to do so, let us suppose
$\{m.\al{}\}=\{last(N).\al{}\}$. \\
Then $\{m.\al{}\}-\{last(N).\al{}\}=0$  therefore 
$\{(m-last(N)).\al{}\}=0$
and as only an integer number has a fractional part equal to zero
we have $(m-last(N)).\al{}=k$, $k\in{}\N{}$ from which we conclude
that $\al{}=\frac{k}{m-last(N)}$ which contradicts $\al{}$ irrationnal.
\end{remark} 
\begin{flushright}
$\boxempty{}$
\end{flushright}
Let us prove now the general case.\\
Let us set for the rest of the proof
$M=first(N)+last(N)$.

\begin{lemma}[Relationship between N and M]
\label{leNM}
$N\leq{}M$.
\end{lemma}

\begin{proof}
\ \\
By reductio ad absurdum. We suppose $M<N$ and we show that, in this
case, the point $first(N)+last(N)$ is situated either before $first$, 
or after $last$, which contradicts their definition. \\
Let us show, therefore, that either 
$\{(first(N)+last(N)).\al{}\}<\{first(N).\al{}\}$ or
$\{(first(N)+last(N)).\al{}\}>\{last(N).\al{}\}$:\\
Let us consider the following cases:
\begin{enumerate}
\item $\{first(N).\al{}\}+\{last(N).\al{}\}<1$: \\
since for $N\geq{}2$  $\{first(N).\al{}\}>0$ we can write that\\
$\{first(N).\al{}\}+\{last(N).\al{}\}>\{last(N).\al{}\}$ thus that \\
$\{(first(N)+last(N)).\al{}\}>\{last(N).\al{}\}$.
\item $\{first(N).\al{}\}+\{last(N).\al{}\}\geq{}1$: \\
in the same way we can write, using the fact that 
$0\leq{}\{\}<1$, that
$\{first(N).\al{}\}+\{last(N).\al{}\}-1<\{first(N).\al{}\}$ 
thus that \\
$\{(first(N)+last(N)).\al{}\}<\{first(N).\al{}\}$.
\end{enumerate}
\begin{flushright}
$\boxempty{}$
\end{flushright}
\end{proof}

\begin{lemma}
\label{firstNM}
$first(N)=first(M)$.
\end{lemma}

\begin{proof}
\ \\
By definition of $first$, we know that for all $a$ and $b$ s.t.
$0<a<N\mbox{ and }0<b<N$ we have that if
$\{a.\al{}\}\leq{}\{b.\al{}\}$ then $a=(first(N))$.\\
Let us take $N=M$ and $a=(first(N))$ and then we have for all $b$
s.t. $0<b<M$ if $\{(first(N)).\al{}\}\leq{}\{b.\al{}\}$ 
then $(first(N))=(first(M))$.\\
Now, it is sufficient to show that
$\{(first(N)).\al{}\}\leq{}\{b.\al{}\}$ $\forall{}b$, 
$0<b<M$:\\
For $0<b<N$ it is the definition of $first$ (lemma \ref{first}).\\
For $N\leq{}b<M$ by the reductio ad absurdum: let us suppose that \linebreak
$\{b.\al{}\}<\{(first(N)).\al{}\}$\\
As $b<M=first(N)+last(N)$ we have immediately that\\
$b-first(N)<last(N)<N$ and $b-last(N)<first(N)<N$ \\
and by definition of $first$ and $last$ that
$\{(b-first(N)).\al{}\}\leq{}\{last(N).\al{}\}$ and
$\{first(N).\al{}\}\leq{}\{(b-last(N)).\al{}\}$. \\
Therefore we have, owing to the hypothesis of contradiction and to 
lemmas \ref{first} and \ref{last} that:\\
$\{b.\al{}\}-\{first(N).\al{}\}+1\leq{}\{last(N).\al{}\}$  and\\
$\{first(N).\al{}\}\leq{}\{b.\al{}\}-\{last(N).\al{}\}+1$ which
implies that \\
$\{last(N).\al{}\}+\{first(N).\al{}\}-\{b.\al{}\}-1=0$ thus that \\
$\{(b-first(N)).\al{}\}=\{last(N).\al{}\}$. But, as shown 
in Remark \ref{irrat} this equality compels $\al{}$ to be rational if
$b-first(N)\neq{}last(N)$ which is the case.
\begin{flushright}
$\boxempty{}$
\end{flushright}
\end{proof}

\begin{lemma}
\label{lastNM}
$last(N)=last(M)$.
\end{lemma}

\begin{proof}
\ \\
Symmetrical proof with the previous.
\begin{flushright}
$\boxempty{}$
\end{flushright}
\end{proof}

\begin{lemma}
\label{afterNM}
For all $n$ s.t. $0<n<N-first(N)$ and $last(N)<n<N$
we have $after(N,n)=after(M,n)$.
\end{lemma}

\begin{proof}
\ \\
We proceed by cases:
\begin{enumerate}
\item Case $0<n<N-first(N)$: \\
Using the irrationality of $\al{}$ (counterpart of remark \ref{irrat}) 
we have that to prove this lemma is equivalent to 
$\{after(N,n).\al{}\}=\{after(M,n).\al{}\}$ which is also
equivalent to  \\
$\{after(N,n).\al{}\}\leq{}\{after(M,n).\al{}\}$ and
$\{after(M,n).\al{}\}\leq{}\{after(N,n).\al{}\}$. \\
Let us proceed by cases and by reductio ad absurdum: 
\begin{itemize} 
\item Case $\{after(N,n).\al{}\}\leq{}\{after(M,n).\al{}\}$:\\
Let us suppose that $\{after(N,n).\al{}\}>\{after(M,n).\al{}\}$.\\
According to lemma \ref{points}, we show immediately the following 
property:\\
$\forall{}N\in{}\N{},\forall{}n,k\in{}Circle$, 
if $\{n.\al{}\}<\{k.\al{}\}$ and if
$\{k.\al{}\}\neq{}\{after(N,n).\al{}\}$ then
$\{after(N,n).\al{}\}<\{k.\al{}\}$. Let us use this property 
with\\
$k=after(M,n)$. We directly get the contradiction on condition that :
\begin{itemize}
\item $n\in{}Circle$ i.e. $0<n<N$; true by case 1.
\item $after(M,n)\in{}Circle$ i.e. $0<after(M,n)<N$ true using 
lemma \ref{inter1}.
\item $\{n.\al{}\}<\{after(M,n).\al{}\}$ by definition of
after (lemma \ref{points} and definition \ref{after}) + lemma 
\ref{inter1} (in order to show that $after(M,n)\neq{}0$).
\item $\{after(M,n).\al{}\}\neq{}\{after(N,n).\al{}\}$ true by
hypothesis of contradiction.
\end{itemize}
\item Case $\{after(M,n).\al{}\}\leq{}\{after(N,n).\al{}\}$:\\
Let us suppose that $\{after(M,n).\al{}\}>\{after(N,n).\al{}\}$.
We use the same property taking $k=after(N,n)$ and $N=M$ (except 
in $k$).
\end{itemize}
\item Case $last(N)<n<N$: the proof is done with the same way.
\end{enumerate}
\begin{flushright}
$\boxempty{}$
\end{flushright}
\end{proof}
For $n$ situated in the third gap, the value of $after(N,n)$ 
is given from the following lemma :

\begin{lemma}
\label{cas3}
For all $n$, $N-first(N)\leq{}n<last(N)$ there does not exist 
\linebreak
$k\in{}Circle$ s.t. 
$\{n.\al{}\}<\{k.\al{}\}<\{(n+first(N)-last(N)).\al{}\}$.
\end{lemma}

\begin{proof}
\ \\
Reduction ad absurdum.\\
Let us suppose there exists one $k\in{}Circle$ 
s.t. \\
$\{n.\al{}\}<\{k.\al{}\}<\{(n+first(N)-last(N)).\al{}\}$.\\ 
This $k$ verifies one of the three following cases (total order on 
the real numbers):
\begin{enumerate}
\item if $\{k.\al{}\}<\{after(M.n).\al{}\}$: \\
then $\{n.\al{}\}<\{k.\al{}\}<\{after(M.n).\al{}\}$ which contradicts
the definition of the function $after$. 
\item if $\{k.\al{}\}=\{after(M.n).\al{}\}$: \\
according to 
lemmas \ref{inter1} and \ref{firstNM} $after(M,n)=n+first(N)$. But, 
as $\al{}$ is irrationnal, we ought to have $k=n+first(N)$ which
contradicts the hypothesises $n<last(N)$ and $k<N$ (using lemma 
\ref{leNM}).
\item if $\{k.\al{}\}>\{after(M.n).\al{}\}$: \\
we use the already 
seen property $\forall{}N\in{}\N{},\forall{}j,
k\in{}Circle$, if $\{j.\al{}\}<\{k.\al{}\}$ and if
$\{k.\al{}\}\neq{}\{after(N,j).\al{}\}$ then 
$\{after(N,j).\al{}\}<\{k.\al{}\}$ with $N=M$, $j=n+first(N)$.\\ 
Then we have, using principally lemma \ref{inter1} that\\
$\{(n+first(N)-last(N)).\al{}\}<\{k.\al{}\}$ which contradicts 
the hypothesis.
\end{enumerate} 
\begin{flushright}
$\boxempty{}$
\end{flushright}
\end{proof}

\subsubsection{Proof of theorem \ref{letheoreme}}
\ \\

Let us suppose that the circle includes $M$ points. Then, we know 
how to show the theorem (lemma \ref{inter1}). Now, It is sufficient
``to remove'' the $M-N$ points which are too many.
\begin{enumerate}
\item if $0\leq{}n<N-first(N)$ then: \\
according to lemma \ref{leNM} we have $0\leq{}n<N-first(N)<last(N)$. 
Using lemmas \ref{firstNM}, \ref{lastNM}, \ref{afterNM} and 
\ref{inter1} we immediately get the result. 
\item if $N-first(N)\leq{}n<last(N)$ then: \\
using lemma \ref{cas3},we show that the $M-N$ points from $N$ do not 
exist and by definition of $after$ (lemma \ref{points} and definition 
\ref{after}) we get the result.
\item if $last\leq{}n<N$ then: \\
as in case 1.
\end{enumerate}
\begin{flushright}
$\boxempty{}$
\end{flushright}

\section*{Conclusion}

The proof given in this paper has been developped from a proof
completely formalized in the system {\sf Coq} \cite{BB+97}.

\subsubsection{The advantages of a formal proof}\

With a mathematical theorem such as this, the interest is twofold:
the first consists in indicating the possible limits of the proof 
assistance system in order to improve it; second, is the emphasizing 
the basic mathematical properties or hypotheses used implicitly
during the demonstration.\\
This proof is based on geometrical intuitions and the demostration
of these intuitions often requires, for example, basic notions about
the fractional parts. Even so, these notions are neither easy to
formalize nor to prove in a system where real numbers are not 
naturally found, unlike other types which can be easily defined
inductively. So, one challenge was to prove this theorem from a 
simple axiomatization of the real numbers. The formulation of real 
numbers used for this will be discussed further.

Throughout this work, we confirmed that the {\sf Coq} proof assistant 
system allows us to work out some purely mathematical proofs. For 
more details, see \cite{MAY99}.

Moreover, it is interesting to notice that the theorem shown is, in some 
sence, stronger than that which was stated initially. Indeed, not only 
do we show that there are at most three different lengths of gaps, but 
we can also give their value and their place on the circle. This modified 
statement is due to \cite{RAV88}.\\

From the proof completely formalized in {\sf Coq}, we can, for
instance, compare this informal proof resulting from the formal proof 
with that of  Tony van Ravenstein.\\

\subsubsection{Properties about $\R{}$}\
\begin{itemize}
\item Two possibilities exist to describe the real numbers: we can 
construct the reals or axiomatize them. We chose an axiomatical 
development for reasons of simplicity and rapidity. We can refer to 
\cite{LAN71} and \cite{LAN51} for constructions from Cauchy's 
sequences or Dedekind's cuts. Most properties of the real numbers 
(commutative field, order, the Archimedian axiom) are first order 
properties. On the other hand, the completeness property is a second 
order property, as it requieres to quantify on the sets of 
real numbers. Instead of this axiom, we can put an infinity 
of first order axioms, according to which any odd degree 
has a root in $\R{}$. Hence, we get the "real closed field" notion.
We thus chose axiomatization at the {\it second order} based 
on the fact that $\R{}$ is a {\bf commutative ordered Archimedian and 
complete field}. For these notions, we based our work on \cite{DIE68} 
and \cite{HAR96}.
\item The formal proof showed us that the axiom of completeness 
of the real numbers was not necessary. Therefore, the statement
and the proof of this theorem are true in all the commutative
ordered and Archimedean fields.\\
Archimedes' axiom could also be replaced by a weaker axiom making 
it possible to define only the fractional part.\\
In the same way, E.Fried and V.T.Sos have given a generalization of 
this theorem for groups \cite{FR+92}. 
\end{itemize}

\subsubsection{The fractional parts.}\

Many intermediate lemmas had to be proved. The formal
proof, for instance, made it possible to identify four lemmas 
concerning the fractional part, which had remained implicit in Tony
van Ravenstein's proof, and are at the heart of the proof.
\begin{itemize}
\item if $\{r1\}+\{r2\}\geq{}1$ then $\{r1+r2\}=\{r1\}+\{r2\}-1$
\item if $\{r1\}+\{r2\}<1$ then $\{r1+r2\}=\{r1\}+\{r2\}$
\item if $\{r1\}\geq{}\{r2\}$ then $\{r1-r2\}=\{r1\}-\{r2\}$
\item if $\{r1\}<\{r2\}$ then $\{r1-r2\}=\{r1\}-\{r2\}+1$
\end{itemize}

\subsubsection{Degenerated cases.}\

The formal proof makes it possible to separate the degenerated cases
such that $N=0$, $N=1$ and $\al{}$ rational, which can be passed over 
in silence during an informal proof.

\subsubsection{$\al{}$ irrational.}\ 

$\al{}$ irrational is hypothesis used by Tony van Ravenstein,
but the formalization shows precisely where this
hypothesis is used (cf remark \ref{irrat}). In particular, if 
$\al{}$ is rational, the points can be mingled, and $after$, for 
example, is not then a function.

\subsubsection{first(N) and first(M), last, after.}\

During Tony van Ravenstein's informal proof, we see that we 
can tolerate an inaccuracy in the dependence of $first$, $last$ 
and $after$ to $N$ or $M$. Although this is not a mistake, the formal 
proof showed the necessity of proving these lemmas, which 
are not trivial (lemmas \ref{firstNM}, \ref{lastNM} and 
\ref{afterNM}). The formal proof makes it possible to say precisely 
where those lemmas are used.

\subsubsection{Use of the classical logic.}\

The formal proof carried out in the system {\sf Coq} - from the 
axiomatization of real numbers as a commutative, ordered, 
archimedian and complet field - is a classical proof seeing that an 
intuitionist reading of the total order involves the decidability of 
the equality of the real numbers, which obviously, is not the case. 
Therefore, we can raise the question of the existence of a 
constructive proof of the three gap theorem.\\
We could probably give an intuitionistic proof for each of the two 
cases, according to whether $\al{}$ is rational or irrational 
because we know exactly the length of the gaps between two points 
of the circle. But, the two cases cannot be treated at the same 
time. Thus in our proof it should be supposed that $\al{}$ is 
rational or not and we do not see, so far how to avoid this 
distinction.\\


\begin{thebibliography}{MAY98}
\bibitem[BB+97]{BB+97}Barras Bruno and al. {\em The Coq Proof
Assistance Reference Manual Version~6.3.1.} INRIA-Rocquencourt, 
May 2000. \\
{\sf http://coq.inria.fr/doc-eng.html}

\bibitem[DIE68]{DIE68}Dieudonn\'e Jean {\em El\'ements d'analyse. 
Vol. 1. Fondements de l'analyse moderne.} Gauthier-Villars Paris
(1968)

\bibitem[FR+92]{FR+92}Fried E. and S\'os V.T. {\em A generalization of
the three-diatance theorem for groups.} Algebra Universalis 29 (1992),
136-149 

\bibitem[HAL65]{HAL65}Halton J.H. {\em The distribution of
the sequence \{$\eta{}\xi{}\}\mbox{ }(n=0,1,2,...)$.} Proc. 
Cambridge Phil. Soc. 71 (1965), 665-670

\bibitem[HAR96]{HAR96}Harrison John Robert. {\em Theorem Proving with
the Real Numbers.} Technical Report number 408, University of 
Cambridge Computer Laboratory, December 1996.

\bibitem[LAN51]{LAN51}Landau Edmund {\em Foundations of analysis.
The arithmetic of whole, rational, irrational and complex numbers.}
Chelsea Publishing Company (1951)

\bibitem[LAN71]{LAN71}Lang Serge {\em Algebra.} Addison-Wesley 
Publishing Company (1971)

\bibitem[MAY99]{MAY99}Mayero Micaela {\em The Three Gap Theorem:
Specification and Proof in {\sf Coq}.} Research Report 3848, INRIA,
December 1999.\\
{\sf ftp://ftp.inria.fr/INRIA/publication/publi-ps-gz/RR/RR-3848.ps.gz}

\bibitem[RAV88]{RAV88}Van Ravenstein Tony {\em The three gap
theorem (Steinhaus conjecture).} J. Austral. Math. Soc. (Series A)
45 (1988), 360-370

\bibitem[SLA67]{SLA67}Slater N.B. {\em Gap and steps for the 
sequence $\eta{}\theta{}$ mod 1.} Proc. Cambridge Phil. Soc. 73 
(1967), 1115-1122

\bibitem[SOS57]{SOS57}S\'os V.T. {\em On the theory of diophantine
approximations.} Acta Math. Acad. Sci. Hungar. 8 (1957), 461-472

\bibitem[SOS58]{SOS58}S\'os V.T. {\em On the distribution mod 1
of the sequence $\eta{}\al{}$.} Ann. Univ. Sci. Budapest. 
E\"otv\"os Sect. Math. 1 (1958), 127-134

\bibitem[SWI58]{SWI58}\'Swierckowski S. {\em On successive
settings of an arc on the circumference of a circle.} Fund. Math.
48 (1958), 187-189

\bibitem[SUR58]{SUR58}Sur\'anyi J. {\em \"Uber der Anordnung der
Vielfachen einer reelen Zahl mod 1.} Ann. Univ. Sci. Budapest. 
E\"otv\"os Sect. Math. 1 (1958), 107-111
\end{thebibliography}
\end{document}